# Optical Frequency Comb Calibrated Near Infrared Solar Heterodyne Spectroscopy

Connor Fredrick[1,2,*], Freja Olsen[2,3], Ryan Terrien[3], Suvrath Mahadevan[4],
Franklyn Quinlan[2], Scott Diddams[1,2,†]

[1]Department of Physics, University of Colorado Boulder, 440 UCB Boulder, CO 80309, USA
[2]Time and Frequency Division, National Institute of Standards and Technology, 325 Broadway, Boulder, CO 80305, USA
[3]Department of Physics and Astronomy, Carleton College, Northfield, Minnesota 55057, USA
[4]Department of Astronomy & Astrophysics, The Pennsylvania State University, 525 Davey Lab, University Park, PA 16802, USA
[*]connor.fredrick@colorado.edu or [†]scott.diddams@nist.gov

**Abstract:** We perform heterodyne spectroscopy at 1.56 µm with a tunable laser and thermal radiation from the Sun. The laser tuning is calibrated with a frequency comb, providing a simple spectrometer with absolute frequency tracebility and resolving power of 2,000,000

The frequency of light is the quantity that has been most precisely measured by humans. The best laser spectroscopy has fractional uncertainty at the level of $1 \times 10^{-18}$ [1] and optical frequency combs have enabled measurements with uncertainty at the $10^{-20}$ level [2]. However, for many important systems of spectroscopic interest—such as those in astronomy—the light to be analyzed is thermal in origin and one typically resorts to interferometric (Fourier transform) or diffractive (grating) spectrometers for spectral analysis. In such spectrometers, wavefront errors and technical constraints on size limit the achievable precision. In this regard, heterodyne spectroscopy between a laser and a thermal source remains an intriguing option for measuring the *frequency* instead of the inferred *wavelength* of an astronomical source. Such thermal heterodyne (also called laser heterodyne radiometry, LHR) has been done for many years in the infrared region of the spectrum (e.g. 10 µm) where the scaling of the density of electromagnetic modes given by the Planck Law provides more favorable signal-to-noise ratio [3-5]. However, there are fewer examples of LHR in the near infrared region of the spectrum, most of which have been focused on gas spectroscopy of the earth's atmosphere [6-7].

In this paper, we explore LHR in the 1.5 µm region of the solar spectrum where widely tunable diode lasers and mature laser frequency comb technology can be brought to bear. With a simple fiber-intergrated heterodyne spectrometer having no mechanical parts, we readily achieve spectral resolving power of $\lambda/\Delta\lambda = 2,000,000$, which would otherwise require meter-scale gratings or optical delay. We detect solar iron lines with signal to noise of 500 in 10 minutes and can track the line shape and line center with <20 MHz precision, limited by the relative earth-sun Doppler shift. Our work is immediately relevant to studies of stellar activity and ongoing efforts in exoplanet science to disentangle such activity from center-of-mass Doppler shifts [8]. Furthermore, this lays the groundwork for future extensions in which massively-parallel heterodyne with frequency combs could be employed not only for spectroscopy but for long baseline phased-array imaging [9].

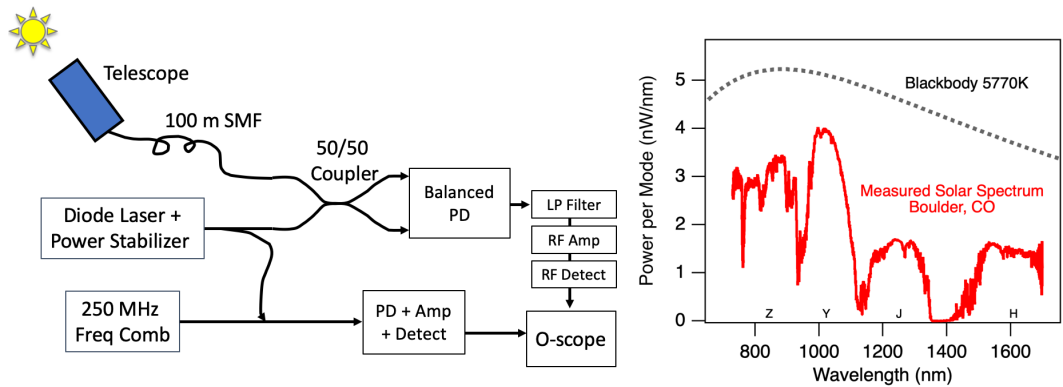

**Fig 1**. (left) Experimental setup. Sunlight is coupled into single mode fiber with a simple telescope. This is subsequently heterodyned with a CW laser in a balanced photodiode (PD). The CW laser frequency is swept in time and its frequency is continuously tracked by a 250 MHz Er:fiber frequency comb. The photodiode signals are processed with common radio frequency (RF) amplifiers and filters before being digitized with and oscilloscope. (right) Spectral density of sunlight coupled into single-mode fiber (SMF28) and measured with a grating spectrometer. Attenuation due to atmospheric water and other gases is clearly seen (e.g. at 1400 nm). The Planck blackbody spectrum for a thermal source at 5770K (which approximates the Sun) is also plotted.

In our experiments (Fig. 1), a single frequency (CW) laser is combined with a thermal source (solar light) in a single-mode optical fiber and the heterodyne signal is generated on a photodetector. Amplitude control of the CW laser and balanced detection minimize technical noise. The sunlight near the laser is mixed down to baseband and the resulting radio frequency power is proportional to the optical power of the source. The resolution of the LHR spectrometer is set by the range of RF frequencies over which the power is integrated. To acquire a spectrum the laser is scanned across the region of interest and the heterodyne power with the sunlight is recorded. With a bandpass in the LHR of 100 MHz we have observed a selection of iron lines in the solar spectrum near 1560 nm with an approximate resolving power of 2,000,000 (see Fig. 2A and 2B). Simultaneously, the CW laser is heterodyned with a frequency comb to calibrate its frequency [10]. The comb provides an equally-spaced grid of optical frequencies with absolute precision traceable to the SI second. The precise reference of the laser frequency comb provides means to track the center frequency of a particular solar line. In this way, we can measure the combined effects of the imprecision of the tracking of our solar telescope and the change in relative velocity between the earth and the Sun. These effects result in a Doppler shift of the particular line (revealed as a frequency drift) that we can read out in real time (Fig. 2C). Analysis of the fluctuations of this frequency drift show an absolute frequency instability of better than 20 MHz in a 20 second averaging window.

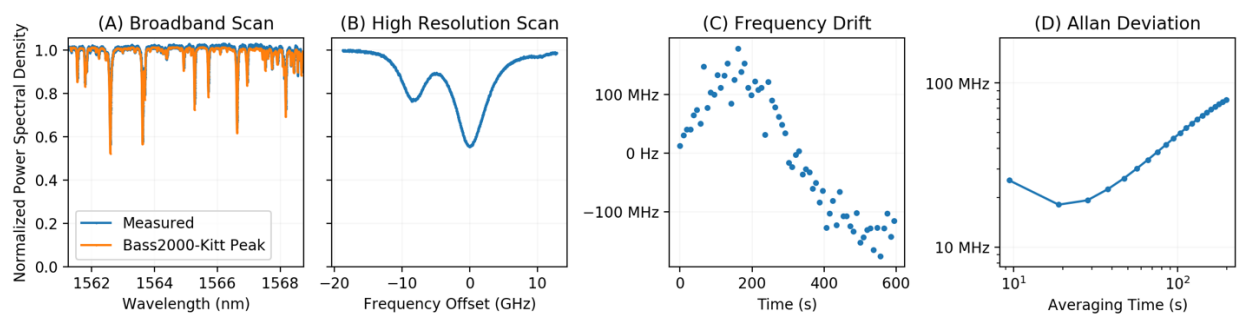

Fig. 2. (A) LHR broad bandwidth measurement of solar lines by scanning the CW laser over 8 nm. Our LHR measurement shows excellent agreement with more conventional high-resolution solar spectrum. (B) High resolution (100 MHz) LHR recording of a solar iron line showing the achievable SNR of 500 in 10 minutes of averaging. (C) Continuous tracking of the center frequency of a solar iron line. Each point is 10 s of averaging. (D) Allan deviation of the data of frame (C), showing measurement precision of <20 MHz in 20 s of averaging.

In summary, using laser heterodyne radiometry with a resolution of 100 MHz we have observed a selection of iron lines in the solar spectrum near 1560 nm with an approximate resolving power of 2,000,000. With a laser frequency comb as the wavelength calibration source we have achieved absolute frequency stability on measurements of single lines at better than 20 MHz in a 20 second averaging window, and SNR of over 500 when averaging up to a 10 minute exposure. Future development will allow detailed spectroscopy of solar lines to understand variability, stellar activity and help pave a path to mitigating these effects in the search for Earth analogs around Sun-like stars.